\documentclass[epj]{svjour}

\usepackage[utf8x]{inputenc}
\usepackage{ucs}
\usepackage{subfig}
\usepackage{graphicx}
\usepackage{tabularx}
\usepackage{float}
\usepackage{latexsym}
\usepackage{threeparttable}
\usepackage{standalone}
\usepackage[normalem]{ulem}
\usepackage{mathtools}
\usepackage{booktabs, dcolumn}
\usepackage{color}
\usepackage{latexsym, amsfonts, amssymb,
txfonts, pxfonts, wasysym}
\newtheorem{thm}{Theorem}

\begin{document}
\title{Rényi entropies of the highly-excited states of multidimensional harmonic oscillators by use of strong Laguerre asymptotics}
\author{A.I. Aptekarev\inst{1}, D.N. Tulyakov\inst{1}, I.V. Toranzo\inst{2,3} \and J.S. Dehesa\inst{2,3}% etc
\thanks{\emph{e-mail:} dehesa@ugr.es}%
}                     % Do not remove
\institute{M.V. Keldysh Institute for Applied Mathematics, Russian Academy of Sciences, Moscow, Russia \and Departamento de F\'{\i}sica At\'{o}mica, Molecular y Nuclear, Universidad de Granada, Granada 18071, Spain \and Instituto Carlos I de F\'{\i}sica Te\'orica y Computacional, Universidad de Granada, Granada 18071, Spain}
\date{Received: date / Revised version: date}
\abstract{
The Rényi entropies $R_{p}[\rho]$, $p>0,\neq 1$ of the highly-excited quantum states of the $D$-dimensional isotropic harmonic oscillator are analytically determined by use of the strong asymptotics of the orthogonal polynomials which control the wavefunctions of these states, the Laguerre polynomials. This Rydberg energetic region is where the transition from classical to quantum correspondence takes place. We first realize that these entropies are closely connected to the entropic moments of the quantum-mechanical probability $\rho_n(\vec{r})$ density of the Rydberg wavefunctions $\Psi_{n,l,\{\mu\}}(\vec{r})$; so, to the $\mathcal{L}_{p}$-norms of the associated Laguerre polynomials. Then, we determine the asymptotics $n\to\infty$ of these norms by use of modern techniques of approximation theory based on the strong Laguerre asymptotics. Finally, we determine the dominant term of the Rényi entropies of the Rydberg states explicitly in terms of the hyperquantum numbers ($n,l$), the parameter order $p$ and the universe dimensionality $D$ for all possible cases $D\ge 1$. We find that (a) the Rényi entropy power decreases monotonically as the order $p$ is increasing and (b) the disequilibrium (closely related to the second order Rényi entropy), which quantifies the separation of the electron distribution from equiprobability, has a quasi-Gaussian behavior in terms of $D$.
} %end of abstract

\maketitle
\section{Introduction}
\label{intro}
Harmonicity is one of the most frequent and useful approximations to simplify and solve the Schrödinger equation of the physical many-body systems. It often provides a deeper quantitative insight into the physical system under investigation, and in many cases allows for the conceptual understanding of physics in a straightforward and intuitive way. Moreover, the solutions of the wave equations of complex physical systems within this approximation are very valuable tools for checking and improving complicated numerical methods used to study such systems.\\

The one-dimensional isotropic harmonic oscillator first and then the $D$-dimensional ($D>1$) oscillator, have been widely used through the history of physics since the 1926-dated seminal paper of Heisenberg \cite{heisenberg}. Indeed they have been used in a great diversity of fields from fractional and quantum statistics \cite{rovenchak,schilling} up to quantum many-body physics \cite{yanez,bouvrie,koscik,carlos,linho1,armstrong1,armstrong2,armstrong3,armstrong4} and black-holes thermodynamics \cite{bombelli,srednicki}, and they have been applied to gain insight into numerous quantum phenomena and systems ranging from heat transport \cite{asadian} and entanglement \cite{eisert,benavides} to Keppler systems \cite{meer}, quantum dots \cite{johnson,koscik,nazmitdinov}, neural networks \cite{agliari}, cold atomic gases \cite{gajda,tempere} and systems with ontological states \cite{hoof}. Let us also remark that the oscillator wavefunctions saturate the various mathematical realizations of the quantum uncertainty principle of Heisenberg and entropic types, which are based on the variance and its moment generalizations (Heisenberg-like uncertainty relations) \cite{sanchez,zozor} and the Shannon entropy \cite{bialynicki1,rudnicki}, Rényi entropy \cite{bialynicki2,vignat} and the Fisher information \cite{romera,sanchez} (entropic uncertainty relations), respectively.\\
 
The spatial extension or spreading of the position probability densities $\rho(\vec{r})$ of a $D$-dimensional isotropic harmonic oscillator, which control all its fundamental properties, has been examined by means of their central moments, particularly the second one (i.e., the variance) \cite{zozor}. It can be complementarily described in the framework of Information Theory by use of the entropic moments of these densities and some related entropic measures \cite{gadre1,yanez_1994,assche_1995,dehesa_1998,ghosh,dehesa_2001}, what is much more adequate because they do not depend on any specific point of their domain of definition, contrary to what happens with the moments about the origin and the central moments. The entropic moments of $\rho(\vec{r})$ are defined as 
\begin{equation}
\label{eq:entropmom}
W_{p}[\rho] = \int_{\mathbb{R}^D} [\rho(\vec{r})]^{p}\, d\vec{r} =\| \rho\|_p^p;\quad p\ge 0, 
\end{equation}
where the  position $\vec{r}  =  (x_1 ,  \ldots  , x_D)$ in hyperspherical units  is  given as $(r,\theta_1,\theta_2,\ldots,\theta_{D-1})      \equiv
(r,\Omega_{D-1})$, $\Omega_{D-1}\in S^{D-1}$, where $r \equiv |\vec{r}| = \sqrt{\sum_{i=1}^D x_i^2}
\in [0  \: ;  \: +\infty)$  and $x_i =  r \left(\prod_{k=1}^{i-1}  \sin \theta_k
\right) \cos \theta_i$ for $1 \le i \le D$
%\sin\theta_1 \ldots \sin\theta_{k-1}  \cos\theta_k$, 
and with $\theta_i \in [0 \: ; \: \pi), i < D-1$, $\theta_{d-1} \equiv \phi \in [0 \: ; \: 2
\pi)$. By  convention $\theta_D =  0$ and the  empty product is the  unity. And the volume element is naturally
\[
d\vec{r} =r^{D-1}drd\Omega_{D} , \quad d\Omega_{D} = \left(\prod_{j=1}^{D-2}\sin^{2\alpha_{j}}\theta_{j}\right)d\phi,
\]
with $2\alpha_{j}= D-j-1$. The symbol $\|\cdot\|_p$     denotes    the    $L_p$     norm    for    functions:
$\|\Phi\|_p=\left(\int_{\mathbb{R}^D} |\Phi(\vec{r})|^p d\vec{r}\right)^{1/p}$. 
The knowledge of the entropic moments or their closely connected quantities, the \textit{Rényi entropies} $R_{p}[\rho]$ (also called by \textit{information generating functionals} in other contexts \cite{golomb}), completely characterize the density $\rho(\vec{r})$. They are defined \cite{renyi1} as
\begin{equation}
\label{eq:renentrop}
R_{p}[\rho] =  \frac{1}{1-p}\ln W_{p}[\rho]; \, 0<p<\infty,\, p \neq 1.
\end{equation}
Note that these quantities include the Shannon entropy (which measures the total extent of the density), $S[\rho] = \lim_{p\rightarrow 1} R_{p}[\rho]$, and the disequilibrium (which quantifies the separation of the density with respect to equiprobability), $\langle\rho\rangle = \exp(R_{2}[\rho])$, as two important particular cases. For a revision of their properties see \cite{aczel,dehesa_88,dehesa_89,romera_01,leonenko,guerrero,jizbad} and the reviews \cite{dehesa_sen12,bialynicki3}. The R\'enyi entropies and their associated uncertainty relations have been widely used to investigate a great deal of quantum-mechanical properties and phenomena of physical systems and processes \cite{bialynicki2,dehesa_sen12,bialynicki3,jizbad}, ranging from the quantum-classical correspondence \cite{sanchezmoreno} and quantum entanglement \cite{bovino} to pattern formation and Brown processes \cite{cybulski1,cybulski2}, fractality and chaotic systems \cite{back,jizba}, quantum phase transition \cite{calixto} and disordered systems \cite{varga}. Moreover, the knowledge of these quantities allows us to reconstruct the corresponding probability density under certain conditions \cite{romera_01,jizba2}.\\

In this work we will investigate the Rényi entropies of the quantum $D$-dimensional oscillator states of the potential $V_{D}(r) = \frac{1}{2}\lambda^{2}r^{2}$, which are known to be described in position space \cite{yanez_1994,dong} by the eigenfunctions 
\begin{eqnarray}
\label{eq:wavpos}
\Psi_{n,l,\{\mu\}}(\vec{r}) &=& \left[\frac{2n!\lambda^{l+\frac{D}{2}}}{\Gamma(n+l+\frac{D}{2})} \right]^{\frac{1}{2}}r^{l}e^{-\frac{\lambda\,r^{2}}{2}}L^{l+D/2-1}_{n}(\lambda\, r^{2}) \nonumber \\
& & \times \, \mathcal{Y}_{l,\{\mu\}}(\Omega_{D-1}),
\end{eqnarray}
and the corresponding energetic eigenvalues
\begin{equation}
\label{eq:oscillenerg}
E_{n,l}= \lambda\left(2n+l+\frac{D}{2}\right),
\end{equation}
where $n=0,1,2,\ldots$ and $l=0,1,2,\ldots$. The symbol $L^{\alpha}_{n}(t)$ denotes the Laguerre polynomial of paramater $\alpha$ and degree $n$ (see definition in Eq. (18) below), and $\mathcal{Y}_{l,\{\mu\}}(\Omega_{D})$ represents the hyperspherical harmonics defined by
\begin{eqnarray}
\label{eq:hyperspherarm}
\mathcal{Y}_{l,\{\mu\}}(\Omega_{D-1}) &=& \mathcal{N}_{l,\{\mu\}}e^{im\phi}\nonumber\\
& & \times \prod_{j=1}^{D-2}C^{\alpha_{j}+\mu_{j+1}}_{\mu_{j}-\mu_{j+1}}(\cos\theta_{j})(\sin\theta_{j})^{\mu_{j+1}}\nonumber\\
\end{eqnarray}
with the normalization constant
\begin{equation}
\label{eq:normhypersphar}
\mathcal{N}_{l,\{\mu\}}^{2} = \frac{1}{2\pi}\times\nonumber\\
\prod_{j=1}^{D-2} \frac{(\alpha_{j}+\mu_{j})(\mu_{j}-\mu_{j+1})![\Gamma(\alpha_{j}+\mu_{j+1})]^{2}}{\pi \, 2^{1-2\alpha_{j}-2\mu_{j+1}}\Gamma(2\alpha_{j}+\mu_{j}+\mu_{j+1})},\nonumber\\
\end{equation}
where the orbital quantum number $l$ and the magnetic quantum numbers $\{\mu\}$ are integers satisfying
\[
l\geq \mu_{1} \geq \mu_{2} \geq \ldots \geq |\mu_{D-1}| \equiv |m|,
\]
and the symbol $C^{\lambda}_{n}(t)$ denotes the Gegenbauer polynomial of degree $n$ and parameter $\lambda$. Atomic units are used throughout the paper.\\

Then, the position probability density of the $D$-dimensional isotropic harmonic oscillator is given by the the squared modulus of the position eigenfunction as follows 
\begin{eqnarray}
\label{eq:denspos}
\rho(\vec{r}) &=& | \Psi_{n,l,\{\mu\}}(\vec{r}) |^2\nonumber\\ &=& \frac{2n!\lambda^{l+\frac{D}{2}}}{\Gamma(n+l+\frac{D}{2})}r^{2l}e^{-\lambda\, r^{2}}\left[L^{(l+D/2-1)}_{n}(\lambda\, r^{2})\right]^{2}\nonumber\\ & & \times|\mathcal{Y}_{l,\{\mu\}}(\Omega_{D-1})|^{2} \nonumber \\
&=& \frac{2n!\lambda^{\frac{D}{2}}}{\Gamma(n+l+\frac{D}{2})}x^{1-\frac{D}{2}}\omega_{l+\frac{D}{2}-1}(x)\left[L^{(l+D/2-1)}_{n}(x)\right]^{2}\nonumber\\
& & \times |\mathcal{Y}_{l,\{\mu\}}(\Omega_{D-1})|^{2}\nonumber \\
&=& 2\,\lambda^{\frac{D}{2}}x^{1-\frac{D}{2}}\omega_{l+\frac{D}{2}-1}(x)[\widehat{L}_{n}^{(l+D/2-1)}(x)]^{2}\nonumber \\
& & \times |\mathcal{Y}_{l,\{\mu\}}(\Omega_{D-1})|^{2} 
\end{eqnarray}
where $x=\lambda\,r^{2}$ and
\begin{equation} \label{eq:c1.1}
\omega_{\alpha}(x) =x^{\alpha}e^{-x}, \, \alpha=l+\frac{D}{2}-1,	
\end{equation}
  is the weight function of the orthogonal and orthonormal Laguerre polynomials of degree $n$ and parameter $\alpha$, here denoted by $L_{n}^{\alpha}(x)$ and $\widehat{L}_{n}^{\alpha}(x)$, respectively. Moreover, it is known \cite{yanez_1994} that the probability density in momentum space (i.e., the squared modulus of the Fourier transform of the position eigenfunction) is given by $\gamma(\vec{p}) =\frac{1}{\lambda^{D}}\rho\left(\frac{\vec{p}}{\lambda}\right)$.\\

Then, by keeping in mind Eqs. (\ref{eq:entropmom}) - (\ref{eq:renentrop}), the main problem in this work is to calculate the quantities
\begin{eqnarray}
\label{eq:entropmom2}
W_{p}[\rho] &=& \int_{\mathbb{R}^D} [\rho(\vec{r})]^{p}\, d\,\vec{r}\nonumber\\ &=& \int\limits_{0}^{\infty}[\rho_{n,l}(r)]^{p}\,r^{D-1}dr
\end{eqnarray}
where we have used the unity normalization of the hyperspherical harmonics
\begin{equation}
	\int_{\mathbb{S}^{D-1}} |\mathcal{Y}_{l,\{\mu\}}(\Omega_{D-1})|^{2}\, d\Omega_{D-1}= 1 \nonumber
\end{equation}
and the radial density function $\rho_{n,l}(r)$ 
\begin{equation}
\label{eq:rhonl}
\rho_{n,l}(X) = 2\,\lambda^{\frac{D}{2}}x^{1-\frac{D}{2}}\omega_{l+\frac{D}{2}-1}(x)[\widehat{L}_{n}^{(l+D/2-1)}(x)]^{2}
\end{equation}

For the low-energy quantum oscillator states (i.e., for low values of the principal quantum number $n$), the analytical expressions of the associated Laguerre polynomials are tractable and the corresponding entropic moments $W_{p}[\rho_{n,l}]$ can be numerically calculated by various accesible quadrature formulas in an effective and sufficiently accurate way. Then, it remains the truly dificult problem: the evaluation of the asymptotics of the quantities
\begin{equation}
\label{eq:rhonl2}
\int\limits_{0}^{\infty}\rho_{n,l}^{p}(r)\,r^{D-1}dr\;,\quad n\to\infty\,,
\end{equation}
which represent the entropic moments of the Rydberg (high-energy) oscillator states. This is the purpose of the present work: to solve this problem in a fully analytical way. Thus, by looking at the expressions (\ref{eq:rhonl}) and (\ref{eq:rhonl2}), this problem converts into an important issue, not yet solved, of the modern Approximation Theory: to study the asymptotics ($n\to\infty$) of the $L_{p}$-norm of the Laguerre polynomials
\begin{equation}\label{eq:c1.2}
N_{n,l}(D,p)=\int\limits_{0}^{\infty}\left(\left[\widehat{L}_{n}^{(\alpha)}(x)\right]^{2}\,w_{\alpha}(x)\right)^{p}\,x^{\beta}\,dx\;,\quad p>0\,,
\end{equation}
where
\begin{equation}\label{eq:c1.3}
\alpha=l+\frac{D}{2}-1\,,\;l=0,1,2,\ldots,\quad\mbox{and}\quad \beta=(p-1)(1-D/2)\;.
\end{equation}
We note that (\ref{eq:c1.1}) and (\ref{eq:c1.3}) guarantee the convergence of
integral (\ref{eq:c1.2}) at zero; i.e. the condition
$$
\beta+p\alpha=pl+\frac{D}{2}-1 > -1\;,
$$
 is always satisfied for physically meaningfull values of the parameters (\ref{eq:c1.3}).\\

\section{Asymptotics of $\mathcal{L}_{p}$ norms of Laguerre polynomials}
\label{sec:2}

In this section we will determine the asymptotics ($n\to\infty$) of the integral functionals $N_{n,l}(D,p)$ of the  (orthonormal) Laguerre polynomials $\widehat{L}_{n}^{(\alpha)}(x)$ defined by Eq. (\ref{eq:c1.2}). It essentially depends on the values of the parameters $D$ and $p$ (i.e. $\alpha,\beta$ and $p$) given by Eq.(\ref{eq:c1.3}).\\

First of all we will make some general comments about the different regions of integration, pointing out the various asymptotical regimes of the Laguerre polynomials and the corresponding dominant contribution. Then, we give the asymptotical results of $N_{n,l}(D,p)$ for all the possible pairs $(D,p)$ in the form of three theorems. Finally we give a detailed proof of these theorems. \\

In fact, to make the ($0, \infty$)-integration in (\ref{eq:c1.2}) for the different values (\ref{eq:c1.3}) of the parameters we have various regions where the Laguerre polynomials have a precise asymptotical representation. First, in the neighborhood of zero (i.e. the left end point of the interval of orthogonality) the Laguerre polynomials can asymptotically be represented by means of Bessel functions as it is pointed out below. Then, to the right, in the bulk region of zeros location, the oscillatory behavior
of the polynomials is modelled asymptotically by means of the trigonometric functions; and at the neighborhood
of the extreme right zeros, asymptotics of the polynomials is given by Airy functions. Finally in the neighborhood of the infinity point, the polynomials has  growing asymptotics. Moreover there are regions where these asymptotics match each
other. Namely, asymptotics of the Bessel functions for large arguments match the trigonometric function, as well as asymptotics of the Airy functions do the same. Altogether there are five asymptotical regimes which can give (depending on $D$ and $p$) the dominant contribution in the asymptotics of $N_{n,l}(D,p)$ . Three of them exhibit the growth of $N_{n,l}(D,p)$ with $n$ by following a power law with an exponent which depends on $D$ and $p$. We call these regimes as Bessel, Airy and cosine (or oscillatory) regimes. Associated to each of these regimes, there is a characteristic constant whose value (as shown below) is
\begin{equation}\label{4}
C_{B}(\alpha,\beta,p):=2\int\limits_{0}^{\infty}t^{2\beta+1}|J_{\alpha} (2t)|^{2p}\,dt\;.
\end{equation}
for the Bessel regime,
\begin{equation}\label{2.5}
C_{A}(p):=
\int_{-\infty}^{+\infty}
\left[
\frac{2\pi}{\sqrt[3]{2}}\,\, {\rm Ai}^2\left(
-\frac{t \sqrt[3]{2}}{2}
\right) \right]^p dt\,.\end{equation}
for the Airy regime, and 
\begin{equation}\label{6}
C(\beta,p):=\displaystyle\frac{2^{\beta+1}}{\pi^{p+1/2}}\,\displaystyle\frac{\Gamma(\beta+1-p/2)\,\Gamma(1-p/2)\,\Gamma(p+1/2)}
{\Gamma(\beta+2-p)\,\Gamma(1+p)}\;.
\end{equation}
for the cosine regime. The symbols $J_{\alpha}(z)$ and $\text{Ai}(-z)$ denote the known Bessel and Airy functions \cite{abramowitz}, respectively, defined below; see Eqs. \eqref{bes}, \eqref{23} and \eqref{24}.\\
In addition, there are two asymptotical regimes corresponding to the transition regions, cosine-Bessel and cosine-Airy. If these
regimes dominate in integral (\ref{eq:c1.2}), then the asymptotics of $N_n(D,p)$ has a factor $\ln n$ besides the power law in $n$. It is also curious to mention that if these regimes dominate then gamma factors in constant $C(\beta,p)$ in (\ref{6}) for the oscillatory cosine regime explode. For the cosine-Bessel regime it happens for $\beta+1-p/2=0$, and for the cosine-Airy regime it happens for
$1-p/2=0$.

\subsection{Asymptotics of the Laguerre polynomials}

Let us now give the asymptotical representation for the
Laguerre polynomials $L_{n}^{(\alpha)}(x)$ defined by
\begin{equation}\label{15}
L_{n}^{(\alpha)}(x)=\sum\limits_{\nu=0}^{n}\displaystyle\left(\substack{n+\alpha
\\n-\nu }\right)\,\frac{(-x)^{\nu}}{\nu!}
\end{equation}
with the norm
\begin{equation}\label{16}
\|L_{n}^{(\alpha)}\|^{2}=\Gamma(\alpha+1)\,\left(\substack{n+\alpha \\n
}\right)\;.
\end{equation}
For the distinct scales of the variable $x$ with respect to $n$ the Laguerre
polynomials have different asymptotics as indicated above.

For the Bessel regime (i.e. when $x$ is small with respect to $n$) there is Hilb asymptotics (see \cite{szego_75} , eq.(8.22.4)):
\begin{equation}\label{17}
\resizebox{0.47\textwidth}{!}{$e^{-\frac{x}{2}}x^{\alpha/2}L_{n}^{(\alpha)}(x)=\frac{(n+\alpha)!}{n!}\,(N\,x)^{-\alpha/2}J_{\alpha}(2\sqrt{N\,x})+\varepsilon(x,n)\;,$}
\end{equation}
where
\begin{equation}\label{18}
\resizebox{0.47\textwidth}{!}{$N=n+\frac{\alpha+1}{2}\;,\qquad \varepsilon(x,n)=\left\{\begin{array}{ll}
x^{\alpha/2+2}\,\underline{\underline{O}}(n^{\alpha})\,, & 0<x<\frac{c}{n} \\
\\
x^{5/4}\,\underline{\underline{O}}(n^{\alpha/2-3/4})\,, & \frac{c}{n}<x<C \\
\end{array}\right.\;,$}
\end{equation}
and the Bessel function is defined by
\begin{equation} \label{bes}
J_{\alpha}(z)=\sum_{\nu=0}^{\infty}\frac{(-1)^{\nu}}{\nu!\,\Gamma(\nu+\alpha+1)}\,\left(\frac{z}{2}\right)^{\alpha+2\nu}\;.
\end{equation}

For the transition region between Bessel regime and oscillatory regime we use
the asymptotics of the Bessel function \cite{abramowitz}:
\begin{equation}\label{19}
\resizebox{0.5\textwidth}{!}{$J_{\alpha}(z)=\displaystyle\sqrt{\frac{2}{\pi
z}}\,\cos\left(z-\frac{\alpha\pi}{2}-\frac{\pi}{4}\right)+e^{|\mbox{Im}\,z|}\underline{\underline{O}}
\left(\frac{1}{z}\right)\;,\quad |\arg z|<\pi\;.$}
\end{equation}

The regimes of oscillatory, growing and Airy types are described by the Plancherel-Rotach asymptotics \cite{plancherel_cmh29,szego_75,tulyakov_rassm10}:
\begin{itemize}
	\item For
$x=(4n+2\alpha+2)\,\cos^{2}\varphi,\;\varepsilon\leqslant\varphi\leqslant\displaystyle\frac{\pi}{2}-\varepsilon
n^{-1/2}$
\begin{equation}\label{20}
\begin{array}{l}
e^{-x/2}\,L_{n}^{(\alpha)}(x)=(-1)^{n}\,(\pi\sin \varphi)^{-1/2}\,x^{-\alpha/2-1/4}\times \\
\\ \quad\qquad \times n^{\alpha/2-1/4} \Big\{\sin \left[\left(n+\frac{\alpha+1}{2}\right)\,(\sin 2\varphi-2\varphi)+\frac{3\pi}{4}\right]\\
\\ \quad\qquad +\,(nx)^{-1/2}\,O(1)\Big\}\;
\end{array}
\end{equation}
\item For
$x=(4n+2\alpha+2)\,\mbox{ch}^{2}\varphi,\;\varepsilon\leqslant\varphi\leqslant\omega$
\begin{equation}\label{21}
\begin{array}{l}
e^{-x/2}\,L_{n}^{(\alpha)}(x)=\frac{1}{2}(-1)^{n}\,(\pi\,\mbox{sinh} \varphi)^{-1/2}\,x^{-\alpha/2-1/4}\times \\
\\
\quad \times n^{\alpha/2-1/4}
\exp\left[\left(n+\frac{\alpha+1}{2}\right)\,(2\varphi-\,\mbox{sinh}
2\varphi)\right]\,[1+O(n^{-1})]\;
\end{array}
\end{equation}
\item And for $x=4n+2\alpha+2-2\left(\frac{2n}{3}\right)^{1/3}\,t\,,\;|t|<const$
\begin{eqnarray}\label{22}
e^{-x/2}\,L_{n}^{(\alpha)}(x)=(-1)^{n}\,\pi^{-1}\,2^{-\alpha-1/3}\,3^{1/3} \times \nonumber\\
 \quad \times n^{-1/3}\{A(t)+O(n^{-2/3})\}
\end{eqnarray}
where the Airy function $A(t)$ 
\begin{equation}\label{23}
A(t)=\frac{\pi}{3}\left(\frac{t}{3}\right)^{1/2}\left[J_{-1/3}\left(2\left(\frac{t}{3}\right)^{\frac{3}{2}}\right)+J_{1/3}\left(2\left(\frac{t}{3}\right)^{\frac{3}{2}}\right)\right]\,
\end{equation}
is the solution of the equation
$$
\frac{d^{2}}{dt^{2}}y+\frac{1}{3}\,t\,y=0\;,
$$
bounded when $t\to\infty$. In (\ref{2.5}) we use normalization for the Airy function as
\begin{equation}\label{24}
A(t)=\frac{\pi}{\sqrt[3]{3}}\,\text{Ai}\left(-t/3\sqrt{3}\right)\;.
\end{equation}
\end{itemize}
During the last two decades there was an essential progress in proving global asymptotical representations for orthogonal polynomials (see Deift et al \cite{deift_93,deift_cpam99,deift_99}, Wong et al \cite{dai,ou} and others \cite{tulyakov_rassm10,AT12}). In practice it means that classical asymptotics formulas (like Hilb and Plancherel-Rotach) hold true in wider domains providing matching of the asymptotics in the transition zones (for example, see in \cite{tulyakov_rassm10} for Hermite polynomials). In our paper we assume that matching of the classical asymptotics holds true for Laguerre polynomials as well.

\subsection{Main results}

Now we are going to state our main asymptotics results. We split them in three theorems.

\begin{thm}\label{T1}
 Let $D\in(2,\infty)$. The weigthed $\mathcal{L}_{p}$-norms of Laguerre polynomials $N_{n,l}(D,p)$, given by (\ref{eq:c1.2}), have the following asymptotical ($n\to\infty$) values:
%Denoting
%\begin{equation}\label{7}
%p^{*}:=\frac{D}{D-1}\;,
%\end{equation}
\begin{equation}\label{8}
\resizebox{0.5\textwidth}{!}{$N_{n,l}(D,p)=\left\{
\begin{array}{ll}
C(\beta,p)\,(2n)^{(1-p)\,D/2}\,(1+\bar{\bar{o}}(1)),\quad & p\in(0,p^{*})\\
\\
\displaystyle\frac{2}{\pi^{p+1/2}n^{p/2}}\,\displaystyle\frac{\Gamma(p+1/2)}{\Gamma(p+1)}\,(\ln n+\underline{\underline{O}}(1))\,,\quad & p=p^{*}\\
\\
C_{B}(\alpha,\beta,p)\,n^{(p-1)D/2-p}\,(1+\bar{\bar{o}}(1)),\quad & p>p^{*}
\end{array}
\right.\;,$}
\end{equation}
where $p^{*}:=\frac{D}{D-1}$, the constants $C$ and $C_{B}$ are defined in (\ref{6}), (\ref{4}) respectively, and the parameters $\alpha \equiv\alpha(l,D)$ and $\beta \equiv \beta(p,D)$ are given by (\ref{eq:c1.3}).
\end{thm}
\textit{Comments}: Let us note that
$$
\resizebox{0.48\textwidth}{!}{$\beta(p^{*},D)-\frac{p^{*}}{2}=(p^{*}-1)\,\left(1-\frac{D}{2}\right)-\frac{p^{*}}{2}=\frac{1}{D-1}\left(1-\frac{D}{2}-\frac{D}{2}\right)=-1\,$,}
$$
so that from (\ref{6}) we have $C(\beta,p)=\infty$. Thus, when $D>2$ we have: for $p\in(0,p^{*})$ the region of $\mathbb{R}_{+}$ where the Laguerre polynomials exhibit the cosine asymptotics contributes with the dominant part in the integral (\ref{eq:c1.2}). For $p=p^{*}$ the transition cosine-Bessel regime
determines the asymptotics of $N_{n,l}(D,p^{*})$, and for $p>p^{*}$ the Bessel regime plays the main role.\\

Let us also highlight that the $\mathcal{L}_{p}$-norm is constant (i.e., independent of $n$) and equal to $C_{B}(\alpha,\beta,p)$, only when $(p-1)D/2-p = 0$. This means that the constancy occurs either when $D=\frac{2p}{p-1}$ or $p=\frac{D}{D-2}$.\\

The next result is

\begin{thm}\label{T2} Let $D=2$. The weigthed $\mathcal{L}_{p}$-norms of Laguerre polynomials $N_{n,l}(D,p)$, given by (\ref{eq:c1.2}), have the following asymptotical ($n\to\infty$) values:
\begin{equation}\label{9}
N_{n,l}(D,p)=\left\{
\begin{array}{ll}
C(0,p)\,(2n)^{(1-p)}\,(1+\bar{\bar{o}}(1))\;, & p\in(0,2)\\
\\
\displaystyle\frac{\ln n+\underline{\underline{O}}(1)}{\pi^{2}n}\;, & p=2\\
\\
\displaystyle\frac{C_{B}(\alpha,0,p)}{n}\,(1+\bar{\bar{o}}(1))\;, & p>2
\end{array}
\right.\;.
\end{equation}
where the constants $C$ and $C_{B}$ are defined in (\ref{6}), (\ref{4}) respectively, and the parameter $\alpha \equiv\alpha(l,D)$ is given by (\ref{eq:c1.3}).
\end{thm}
\textit{Comments}: A peculiarity of the case $D=2$ is the following. We have from Theorems \ref{T1} and \ref{T2}
$$
\lim\limits_{D\to 2+} N_n(D,p)=N_n(2,p)\;,\quad p\in(0,2)\cup(2,\infty)\;.
$$
However, from Theorem~\ref{T1} we have
\begin{equation}\label{10}
\lim\limits_{D\to 2+} N_n(D,2)=\frac{3(\ln
n+\underline{\underline{O}}(1))}{4\pi^{2}n}\;.
\end{equation}
On the other hand, Theorem \ref{T2} states:
$$
N_n(2,2)=\frac{\ln n+\underline{\underline{O}}(1)}{\pi^{2}n}\;.
$$
Indeed, as we shall prove it below, the magnitude of integral $N_n(2,2)$ is performed mainly by two regions of $\mathbb{R}_{+}$ (with the same order of contribution). The first one is at the origin (Bessel-cosine regime), and the second one is around the right-extreme zeros of the Laguerre polynomials (Airy-cosine regime). The first region gives the contribution in $N_n(2,2)$ as in (\ref{10}). The second one gives the rest of the contribution
\begin{equation}\label{11}
\frac{\ln n+\underline{\underline{O}}(1)}{4\pi^{2}n}\;.
\end{equation}
Thus for $D=2$ and $p=2$ we have the competition of two transition regimes, namely the Bessel-cosine and Airy-cosine regimes.\\
Let us also highlight that the $\mathcal{L}_{p}$-norm is constant when $p=1$, being its value $C(0,p) = 1$.

The third, final, result on asymptotics of $N_n(D,p)$ (we recall $\beta$ is defined in (\ref{eq:c1.3})) is the following.

\begin{thm}\label{T3} Let $D\in[0,2)$. The weigthed $\mathcal{L}_{p}$-norms of Laguerre polynomials $N_{n,l}(D,p)$, given by (\ref{eq:c1.2}), have the following asymptotical ($n\to\infty$) values:
\begin{itemize}
	\item For $p\in(0,2]$,
	\begin{equation}\label{12}
N_n(D,p)=\left\{
\begin{array}{ll}
C(\beta,p)\,(2n)^{(1-p)\frac{D}{2}}\,(1+\bar{\bar{o}}(1))\;, & p\in(0,2)\\
\\
\displaystyle\frac{\ln n+\underline{\underline{O}}(1)}{\pi^{2}(4n)^{1-\beta}}\;, & p=2\\
\end{array}
\right.\;.
\end{equation}
\item For $p>2$ and $ 4/3<D<2$,
\begin{equation}\label{14}
\resizebox{0.47\textwidth}{!}{$N_{n,l}(D,p)=\left\{
\begin{array}{ll}
\displaystyle\frac{C_{A}(p)}{\pi^p}\,(4n)^{(\frac{1-2p}{3}+\beta)}(1+\bar{\bar{o}}(1))\,, & p\in(2,\widetilde{p})\\
\\
\left(\displaystyle\frac{C_{A}(p)}{\pi^p}\,4^{(\frac{1-2p}{3}+\beta)}+C_{B}(\alpha,\beta,p)\right)\,n^{-\beta-1}\;,
& p=\widetilde{p}\\
\\
C_{B}(\alpha,\beta,p)\,n^{-\beta-1}\;, & p\in(\widetilde{p},\infty)
\end{array}
\right.,$}
\end{equation}
where $\widetilde{p}:=\displaystyle\frac{-2+3D}{-4+3D}$, and
\item For $p>2$ and $D\leqslant 4/3$,  
\begin{equation}\label{13}
\resizebox{0.47\textwidth}{!}{$N_n(D,p)=\frac{C_{A}(p)}{\pi^
p}\,(4n)^{(\frac{1-2p}{3}+\beta)}(1+\bar{\bar{o}}(1))\,,\quad \quad \quad
p\in(2,\infty)\;.$}
\end{equation}
\end{itemize}
where the constants $C$, $C_{A}$ and $C_{B}$ are defined in (\ref{6}), (\ref{2.5}) and (\ref{4}) respectively, and the parameters $\alpha \equiv\alpha(l,D)$ and $\beta \equiv \beta(p,D)$ are given by (\ref{eq:c1.3}).
\end{thm}
\textit{Comment}: Here we see, that the oscillatory regime in \eqref{12} for $ p\in(0,2)$ matches the same regime in \eqref{8} and \eqref{9} for $ p<p^{*}$. But for $ p=2$ the Airy-cosine regime wins versus Bessel-cosine regime and we have only contribution of \eqref{9} in $N_n(D,p)$. For $p \ge 2$ we get a new phenomena: the role of the oscillatory regime disappears and for the first time  the  Airy and Bessel regimes becomes competitive.\\

Here, the limits as $n\to\infty$ of the $\mathcal{L}_{p}$-norm is constant when $p=1$, $\beta=-\frac{1-2p}{3}$ (i.e. when $p=1+\frac{2}{2-3D}$ or $D=\frac{2}{3}\frac{p-2}{p-1}$), and $\beta=-1$ (i.e. when $p=\frac{D}{D-2}$ or $D=\frac{2p}{p-1}$).

\subsection{Proofs}

For all three theorems we use the unified approach. We split the domain of integration $\mathbb{R}_{+}$ of (\ref{eq:c1.2}) into nine intervals as
\begin{eqnarray}
N_{n,l}(D,p)&=& \displaystyle\frac{\int\limits_{0}^{\infty}((L_{n}^{(\alpha)}(x))^{2}\,w(x))^{p}\,x^{\beta}\,dx}{\|L_{n}^{(\alpha)}\|^{2p}}\nonumber\\
 &=& n^{-p\alpha}\,\left(\sum\limits_{j=1}^{9}I_{j}\right),	
\end{eqnarray}

where
\begin{equation}\label{25}
I_{j}:=\int\limits_{\triangle_{j}}((L_{n}^{(\alpha)}(x))^{2}\,w(x)^{p}\,x^{\beta}\,dx\;,
\end{equation}
and
{\footnotesize\begin{equation}\label{26}
\begin{array}{ll}
  \Delta_{1}=[0,M/n]\,; \qquad
  \Delta_{2}=[M/n,1]\,; \,\,
  &\Delta_{3}=[1,(4-\varepsilon)n]\,;\\ \\
  \Delta_{4}=[(4-\varepsilon)n,4n-n^{\frac13+\theta}]\,; \,\,
  &\Delta_{5}=[4n-n^{\frac13+\theta},4n-Mn^{\frac13}];\\\\
  \Delta_{6}=[4n-Mn^{\frac13},4n]\,; \qquad
  &\Delta_{7}=[4n,4n+Mn^{\frac13}]\,; \\ \\
  \Delta_{8}=[4n+Mn^{\frac13},4n+n^{\frac13+\theta}]\,; \qquad
  &\Delta_{9}=[4n+n^{\frac13+\theta},\infty]\,,
\end{array}
\end{equation}
}
for some big $M>0$, small $\varepsilon>0$ and $\theta>0$. Then we
replace $L_{n}^{(\alpha)}\,w$ in (\ref{25}) by their
asymptotics. For $j=1$ we use Hilb asymptotics
(\ref{17})-(\ref{18}); for $j=2$ we use Hilb asymptotics
(\ref{17})-(\ref{18}) and Bessel function asymptotics (\ref{19});
for $j=3,4$ we use oscillatory asymptotics of Plancherel-Rotach
(\ref{20}); for $j=5,6,7,8$ we use Airy asymptotics of
Plancherel-Rotach (\ref{22}); and for $j=9$ we use growing asymptotics
of Plancherel-Rotach (\ref{21}).

Eventually we estimate the contribution of each integral from
$\{I_{j}\}_{j=1}^{9}$ finding the dominating terms.

\subsection{Proof of Theorem~\ref{T1}}\label{4.1}

Here we have $D>2$ and $p^{*}=\frac{D}{D-1}$.

%\medskip

Let us start with the case $p > p^{*}$. Then in the representation
(\ref{25})-(\ref{26}) for $N_{n,l}(D,p)$ by the sum of integrals
$\sum\limits_{j=1}^{9}I_{j}$, the main contribution for this case is given by $I_{1}$. We have
\begin{eqnarray}
I_{1} &=&\int\limits_{0}^{M/n}(w^{1/2}(x)\,\widehat{L}_{n}^{(\alpha)}(x))^{2p}\,x^{\beta}\,dx \nonumber\\
&=& \int\limits_{0}^{M/n}\Bigg[\left(\frac{(n+\alpha)!}
{n!}\right)^{2}(Nx)^{-\alpha}J_{\alpha}^{2}(2\sqrt{Nx})\nonumber\\
 & &+\, O\left(x^{\alpha/2+2}n^{\alpha}\right)\Bigg]^{p}  x^{p\,\alpha+\beta}\,dx.
\end{eqnarray}
Making the change of the variable $t:=\sqrt{Nx}$, we continue
\begin{equation}\label{27}
\begin{array}{c}
I_{1}\simeq n^{2p\,\alpha}\cdot N^{-p\,\alpha-\beta-1}\int\limits_{0}^{\sqrt{\frac{MN}{n}}}2t^{2p\,\alpha+2\beta+1}t^{-2p\,\alpha}|J_{\alpha}^{2p}|
\,(2t)\,dt\simeq \\
\\
\simeq n^{p\,\alpha-\beta-1}\int\limits_{0}^{\sqrt{M}}2t^{2\beta+1}|J_{\alpha}^{2p}|\,(2t)\,dt\;.
\end{array}
\end{equation}
The last integral converges at zero. Indeed, the integrand has there the
order of singularity $2p\,\alpha+2\beta+1>-1$
due to (\ref{eq:c1.3}). The order of singularity of the integrand at
infinity is $2\beta+1-p < -1$ due to $p>p^{*}$. Since the parameter $M$ is  arbitrary in our partition of $\mathbb{R}_{+}$ in (\ref{26})), we take $M\to\infty$ and obtain
\begin{equation}\label{28}
n^{-p\,\alpha}I_{1}\simeq n^{-\beta-1}\int\limits_{0}^{\infty}2t^{2\beta+1}|J_{\alpha}|^{2p}(2t)\,dt\;.
\end{equation}
In fact, the contribution in $N_{n,l}$ of the remaining integrals
$I_{j},\,j=2,\ldots,9$ for $D>2,\,p>p^{*}$ is less (we will see it
latter). Thus, due to (\ref{eq:c1.3}) and (\ref{4}), asymptotics (\ref{28})
is the same as in (\ref{8}) for $p>p^{*}$.

Now, let us consider the case $p=p^{*}$. Then, the dominant behavior is coming from the two integrals $I_{2}$ and $I_{3}$. Indeed, we have from (\ref{27}) that
\begin{equation}\label{29}
n^{-p\,\alpha}I_{1}=O\left(\frac{M^{p\,\alpha+\beta+1}}{n^{\beta+1}}\right)+\delta_{n}\;,\quad \delta_{n}=\frac{M^{p\,\alpha+\beta+3}}{n^{\beta+3}}\;.
\end{equation}
We note that from (\ref{eq:c1.3}) we have
\begin{equation}\label{30}
\beta-\frac{p^{*}}{2}=(p^{*}-1)\,\left(1-\frac{D}{2}\right)-\frac{p^{*}}{2}=-1\;.
\end{equation}
Taking into account the asymptotics of the Bessel function
(\ref{19}), we have the following estimation or  $I_{2}$:
\begin{eqnarray}
n^{-p\,\alpha}I_{2}&=&\int\limits_{M/n}^{1}J_{\alpha}^{2p}(2\sqrt{Nx})\,x^{\beta}\,dx+\tilde{\delta}_{n}\nonumber\\
&=& \int\limits_{M/n}^{1}\frac{1}{\pi^{p}(Nx)^{p/2}}\Bigg\{\cos\left(2\sqrt{Nx}-(2\alpha+1)\cdot\frac{\pi}{4}\right)\nonumber\\
& &+
\underline{\underline{O}}\left(\frac{1}{\sqrt{N}}\right)\Bigg\}^{2p}\,x^{\beta}\,dx+\tilde{\delta}_{n}\;.
\end{eqnarray}
Using (\cite{aptekarev_rassm95}, Lemma 2.1) we can continue for
$n\to\infty$ as
$$
n^{-p\,\alpha}I_{2} = \frac{1}{\pi}\int\limits_{0}^{\pi}|\cos\theta|^{2p}\,d\theta\,\int\limits_{M/n}^{1}\frac{x^{-p/2+\beta}\,dx}{\pi^{p}\,N^{p/2}}
\,(1+\bar{\bar{o}}(1))\;.
$$
The first integral is
$$
\int\limits_{0}^{\pi}|\cos\theta|^{2p}\,d\theta=\frac{\sqrt{\pi}\,\Gamma(p+1/2)}{\Gamma(p+1)}\;.
$$
Computing the second integral for $p=p^{*}$ (see (\ref{30})), we
obtain
\begin{equation}\label{31}
n^{-p^{*}\alpha}I_{2} = \frac{\Gamma(p^{*}+1/2)\,(\ln n+\underline{\underline{O}}(1))}{\pi^{p^{*}+1/2}\,\Gamma(p^{*}+1)\,N^{p/2}}\;.
\end{equation}
The Plancherel-Rotach asymptotics (\ref{20}) for
$\varphi=\arccos\sqrt{\frac{x}{4N}}$ can be transformed to
\begin{equation}\label{32}
\resizebox{0.49\textwidth}{!} 
{
$\frac{x^{\alpha}}{n^{\alpha}}\left(e^{x/2}L_{n}^{\alpha}(x)\right)^{2} = \frac{2\sin^{2}\left[\frac{1}{2}\sqrt{x(4N-x)}-
2N\arccos\sqrt{\frac{x}{4N}}+\frac{3\pi}{4}\right]+O\left(\frac{1}{\sqrt{nx}}\right)}{\pi\sqrt{x(4N-x)}}\,.$
}
\end{equation}
Substituting it in $I_{3}$ and using (\cite{aptekarev_rassm95},
Lemma 2.1) we have for $I_{3}$, as $n\to\infty$
$$
n^{-p^{*}\alpha}I_{3}=\int\limits_{1}^{(4-\varepsilon)n}\frac{x^{\alpha p^{*}}}{n^{\alpha p^{*}}}\left(e^{x/2}L_{n}^{(\alpha)}(x)\right)^{2p^{*}}
x^{\beta}dx=
$$
$$
=\left(\frac{2}{\pi\sqrt{4n}}\right)^{p^{*}}\,\frac{1}{\pi}\int\limits_{0}^{\pi}|\sin\theta|^{2p^{*}}d\theta\cdot\int\limits_{1}^{(4-\varepsilon)n}
x^{\beta-p^{*}/2}dx\;.
$$
Thus, $I_{3}$ gives the same contribution in $N_{n,l}(D,p^{*}$ as
$I_{2}$ in (\ref{31})
\begin{equation}\label{33}
n^{-p^{*}\alpha}I_{3} = \frac{\Gamma(p^{*}+1/2)\,(\ln n+\underline{\underline{O}}(1))}{\pi^{p^{*}+1/2}\,\Gamma(p^{*}+1)\,N^{p/2}}\;.
\end{equation}
We see from (\ref{29}) that for $p=p^{*}$ the contribution from $I_{1}$
in $N_{n,l}(D,p^{*})$ is less than that from $I_{2}$ and $I_{3}$. The same
can be shown for the contribution of other integrals. Thus, summing
up (\ref{31}) and (\ref{33}) we arrive at (\ref{8}) for $p=p^{*}$.

It remains to consider the case $p\in(0,p^{*})$. The dominant
contribution here is given by $I_{3}$. Substituting asymptotics (\ref{32}) in $I_{3}$, making the change of variable
$t:=\sqrt{\frac{x}{4n}}$ and using (\cite{aptekarev_rassm95}, Lemma
2.1) we arrive to
\begin{equation*}
\resizebox{.49 \textwidth}{!} 
{
$
N^{-p\,\alpha}I_{3}=\left(\frac{2}{\pi 4n}\right)^{p}\,(2\sqrt{n})^{2\beta+2}\,\frac{1}{\pi}\int\limits_{0}^{\pi}|\sin\theta|^{2p}d\theta\cdot
\int\limits_{0}^{1}\frac{t^{2\beta+1}\,dt}{t^{p}(1-t^{2})^{p/2}}\,(1+\bar{\bar{o}}(1))\;.
$
}
\end{equation*}
The last integral can be evaluated explicitly as
$$
\int\limits_{0}^{1}\frac{t^{2\beta+1}\,dt}{t^{p}(1-t^{2})^{p/2}}=\frac{1}{2}\,\frac{\Gamma(\beta+1-p/2)\,\Gamma(1-p/2)}{\Gamma(\beta+2-p)}\;.
$$
Thus, we obtain
\begin{equation}\label{34}
\resizebox{.49 \textwidth}{!} 
{
$
n^{-p^{*}\alpha}I_{3}=\frac{2^{\beta+1}}{\pi^{p+1}}\,\frac{\Gamma(\beta+1-p/2)\,\Gamma(1-p/2)\,\Gamma(1+p/2)}
{\Gamma(\beta+2-p)\,\Gamma(1+p)}\,(2n)^{1-p+\beta}\,(1+\bar{\bar{o}}(1))\;.
$
}
\end{equation}
It is clear that the contributions of $I_{1}$ and $I_{2}$ are less than
$I_{3}$. The same can be shown for the contribution of other integrals. Theorem is proved.

\subsection{Proof of Theorem~\ref{T2}}\label{4.2}

Here we have $D=2$. Then, $\beta\equiv 0$ and $p^{*}=2$.

\medskip

Let us start with the case $p>2$. As for the case ($D>2$, $p>p^{*}$), according to (\ref{25}) -- (\ref{26}) we can see that the dominant contribution in $N_{n,l}(D,p)$ is given by $I_{1}$. Indeed, we have
$$
\int\limits_{0}^{M/n}\left(w^{1/2}(x)\,\widehat{L}_{n}^{(\alpha)}(x)\right)^{2p}\,dx=
$$
{\footnotesize$$
=\int\limits_{0}^{M/n}\left[\frac{n!}{(n+\alpha)!}\,\left(\frac{(n+\alpha)!}{n!}\right)^{2}(Nx)^{-\alpha}J_{\alpha}^{2}(2\sqrt{Nx})+
x^{\alpha+4}O(n^{\alpha})\right]^{p}x^{p\,\alpha}\,dx  
$$}
$$= \frac{1}{n}\left(\int\limits_{0}^{\sqrt{M}}2t\,|J_{\alpha}|^{2p}(2t)\,dt+ \bar{\bar{o}}(1)\right)\;.
$$
Since $M$ is an arbitrary constant, we let $M\to\infty$. At the same
time, we see that the sum $J_{6}+J_{7}$ also gives a perceptible
contribution
\begin{equation}\label{35}
\resizebox{.49 \textwidth}{!} 
{
$
\int\limits_{4N-Mn^{1/3}}^{4N+Mn^{1/3}}\left(w^{1/2}(x)\,\widehat{L}_{n}^{(\alpha)}(x)\right)^{2p}dx=\int\limits_{-M}^{M}\left[(2n)^{-2/3}A_{i}^{2}
\left(-\frac{t}{2^{4/3}}\right)\right]^{p}n^{1/3}\,dt\,(1+\bar{\bar{o}}(1)\;.
$
}
\end{equation}
However, for $p>2$
\begin{equation}\label{36}
1/3-p\,2/3<-1\;.
\end{equation}
Thus the only contribution of $I_{1}$ plays the role, and we obtain
(\ref{9}) for $p>2$.

Let us now consider the case $p=2$. In comparison with the case ($D>2$, $p=p^{*}$), not only the transition zone for the Bessel-cosine regimes (i.e. integrals $I_{2}$ and $I_{3}$) plays the role, but the transition zone for the cosine-Airy regimes (i.e. integrals $I_{4}$ and $I_{5}$) plays the role too.

For $I_{2}$ and $I_{3}$, substituting $p^{*}=2$ in (\ref{31}) and
(\ref{33}), we get
\begin{equation}\label{37}
n^{-2\alpha}(I_{2}+I_{3})=\frac{3\ln n+\underline{\underline{O}}(1)}{4\pi^{2}n}\;.
\end{equation}
The second transition zone is \\ $\left[(4-\varepsilon)n,
4n-n^{1/3+\theta}\right]\cup\left[4n-n^{1/3+\theta},4n-M\cdot
n^{1/3}\right]$. For the oscillatory Plancherel-Rotach asymptotics
(\ref{20}) we have
\begin{equation}\label{38}
\resizebox{.49 \textwidth}{!} 
{
$
\begin{array}{c}
\int\limits_{(4-\varepsilon)N}^{4N-n^{1/3+\theta}}\left[\frac{2\sin^{2}\left(\frac12\sqrt{x(4N-x)}-2N\arccos\sqrt{\frac{x}{4N}}+\frac{3\pi}{4}
\right)+O\left(\frac{1}{\sqrt{Nx}}\right)}{\pi\sqrt{x(4N-x)}}\right]^{2}\,dx= \\
\\
=\frac{1}{\pi}\int\limits_{0}^{\pi}\sin^{4}\varphi\,d\varphi\cdot\int\limits_{4N}^{4N-n^{1/3+\theta}}\frac{4\,dx}{\pi x(4N-x)}=\frac{3}{8\pi^{2}n}
\left((\frac23-\theta)\ln n+\underline{\underline{O}}(1)\right)\;.
\end{array}
$}
\end{equation}
For $I_{5}$ using (\ref{22}) and asymptotics for the Airy function
(see in \cite{deift_99})
$$
A_{i}^{4}\left(-\frac{t}{2^{4/3}}\right)\simeq \frac{(1+\sin(t^{3/2}/3))^{2}}{4\pi^{2}(t/2^{4/3})}\;,\quad t\to\infty\,,
$$
we obtain
\begin{equation}\label{39}
\resizebox{.49 \textwidth}{!} 
{
$
\begin{array}{c}
\int\limits_{4N-n^{(1/3+\theta)}}^{4n-Mn^{1/3}}(w^{1/2}(x)\widehat{L}_{n}^{(\alpha)}(x))^{2}dx\simeq \int\limits_{M}^{n^{\theta}}\left[
(2n)^{-2/3}A_{i}^{2}\left(-\frac{t}{2^{4/3}}\right)\right]^{2}n^{1/3}dt\simeq \\
\\
\simeq \frac{1}{4\pi^{2}n}\int\limits_{0}^{\pi}(1+\sin\varphi)^{2}d\varphi\int\limits_{M}^{n^{\theta}}\frac{dt}{t}=
\frac{3(\theta\ln n +\underline{\underline{O}}(1))}{8\pi^{2}n}\;.
\end{array}
$}
\end{equation}
Summing (\ref{38}), (\ref{39}) and (\ref{37}), we obtain (\ref{9}) for
$p=2$.

The remaining case is $p<2$. Here we proceed in the same manner as for
the case ($D>2$, $p<p^{*}$), and we obtain (\ref{34}) for $\beta=0$.
Theorem is proved.

\subsection{Proof of Theorem~\ref{T3}}\label{4.3}

Here we have $D\in[0,2),\,\, \beta>0$ for $p>1$; therefore $p^{*}=2$, as
in the previous case.

\medskip

Let us start with the case $p>2$. Now the competition between $I_{1}$ and
$I_{6}+I_{7}$ becomes crucial. We already know for $I_{1}$ from
(\ref{28}) that
$$
n^{-p\,\alpha}I_{1}=C_{B}n^{-\beta-1}\;.
$$
To obtain the asymptotics for $n^{-p\,\alpha}(I_{6}+I_{7})$ we substitute
$x^{\beta}$ in the left-hand side of (\ref{35})
$$
\int\limits_{4n-Mn^{1/3}}^{4n+Mn^{1/3}}(w^{1/2}(x)\widehat{L}_{n}^{(\alpha)}(x))^{2p}x^{\beta}dx\simeq 2^{2\beta}n^{\frac{1-2p}{3}+\beta}C_{A}\;.
$$
Now, instead of inequality (\ref{36}) we have for $D>4/3$ the solution
$p=\tilde{p}$ of the equation (where $\beta$ is from (\ref{eq:c1.3}))
$$
-\beta-1=1-\frac{2p}{3}+\beta\Rightarrow \tilde{p}=\frac{-2+3D}{-4+3D}\;.
$$
Thus, we have obtained (\ref{13}) and (\ref{14}).

Now let us consider that $p=2$. In comparison with the previous cases, we have that the only contribution which plays a role is coming from the transition zone for the cosine-Airy regimes. Substituting $x^{\beta}$ in the left-hand sides of (\ref{38}) and
(\ref{39}) we arrive at (\ref{12}), $p=2$.

Finally for $p\in(0,2)$, we have
$$
1+\beta-p>-\beta-1\;,
$$
and
$$
1+\beta-p>\frac{1-2p}{3}+\beta\;.
$$
Thus, only the oscillatory integral $I_{3}$ gives the contribution
to the asymptotics of $N_{n,l}(D,p)$, and from (\ref{34}) we complete
proof of (\ref{12}).

Theorem is proved.

\newpage

\section{Rényi entropy powers for Rydberg $D$-dimensional oscillator states}

In this section the asymptotical results obtained in the previous section are applied to obtain the Rényi entropies (or better, the Rényi entropy powers, which have position physical units) of the Rydberg states of the multidimensional harmonic oscillator. The Rényi entropy powers, $\mathcal{N}_{p}[\rho]$, of the density $\rho$ is given by
\begin{equation}
\label{eq:renyipow}
\mathcal{N}_{p}[\rho] := e^{R_{p}[\rho]} = \left(\int \rho(x)^{p}\, dx\right)^{\frac{1}{1-p}}.
\end{equation}
Taking into account Eqs. (\ref{eq:renentrop}), (\ref{eq:entropmom2}), (\ref{eq:rhonl}) and (\ref{eq:c1.2}), we obtain the following expressions 
\begin{eqnarray}
\label{eq:renyiasymp}
R_{p}[\rho] &=& \frac{1}{1-p}\, \ln\left[2^{p-1}\lambda^{\frac{D}{2}(p-1)}N_{n,l}(D,p)\right], \nonumber\\
%\mathcal{N}_{p}[\rho] &=& \left[2^{p-1}\lambda^{\frac{D}{2}(p-1)}N_{n,l}(D,p)\right]^{\frac{1}{1-p}}\nonumber\\
\mathcal{N}_{p}[\rho] &=& \frac{1}{2}\lambda^{-\frac{D}{2}}\left[ N_{n,l}(D,p)\right]^{\frac{1}{1-p}}
\end{eqnarray}
for the Rényi entropies and the Rényi entropy powers, respectively, of an arbitrary quantum state of the $D$-dimensional isotropic harmonic oscillator in terms of the $\mathcal{L}_{p}$-norms $N_{n,l}(D,p)$ of the orthonormal Laguerrre polynomials associated to the state wavefunction given by the three previous theorems. The involved parameters within the norms, $\alpha \equiv\alpha(l,D) = l+\frac{D}{2}-1\, (l=0,1,2,\ldots)$ and $\beta \equiv \beta(p,D) = (p-1)(1-D/2),$ are taking from (\ref{eq:c1.3}). Note that these information-theoretic quantities depend on the spatial dimension $D$ as well as on the order parameter $p$ for each pair ($n,l$). In the numerical calculations performed heretoforth we will assume that $\lambda = 1$ without any loose of generality. Atomic units are used everywhere as already pointed out.\\

Let us now discuss these two Rényi-type quantities with  $D\ge 2$ and $q\ne 1$ from Eqs. (\ref{eq:renyiasymp}) in various ways. The Rényi entropies for the Rydberg states of the one-dimensional isotropic harmonic oscillator have been recently studied \cite{ADST} in a monographic way, because the polynomials involved in this case are of Hermite type. The limiting case $p \rightarrow 1$ (Shannon entropy) will be analyzed separately elsewhere for any $D$-dimensional oscillator system.\\

\begin{figure}[H]
 \minipage{0.50\textwidth}
 \label{fig:l0_n50_D2}
 \includegraphics[width=\linewidth]{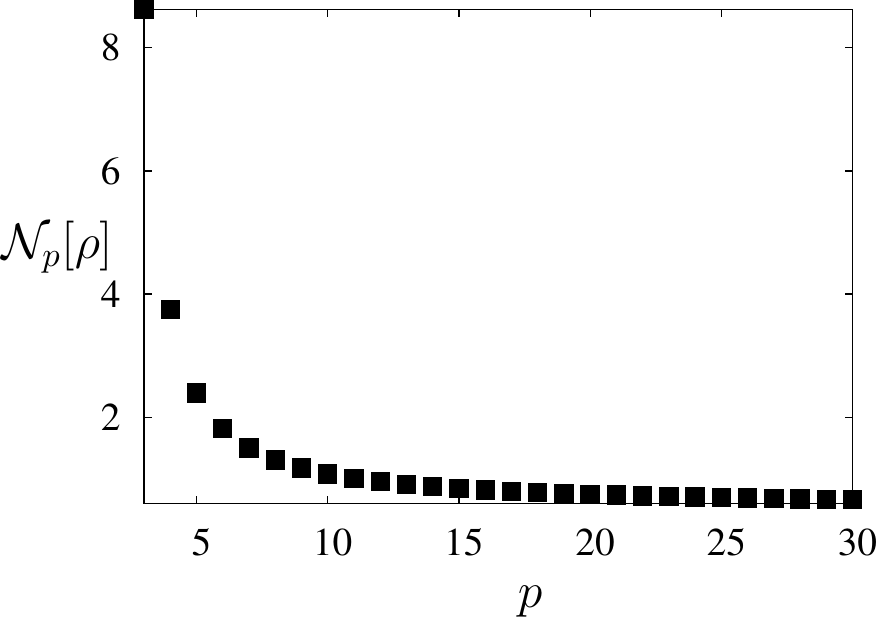}
 \vspace{4mm}
 \endminipage\hfill
 \minipage{0.50\textwidth}
   \includegraphics[width=\linewidth]{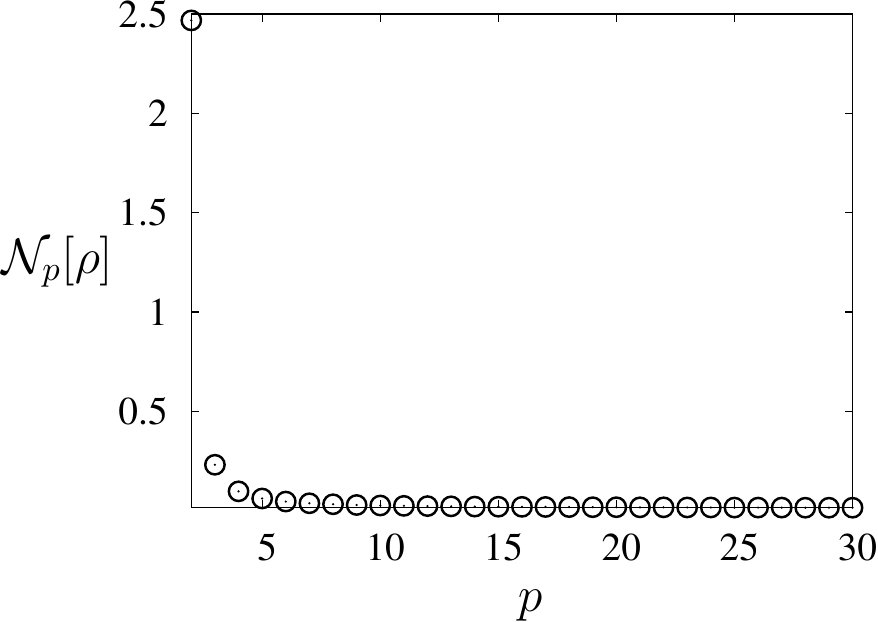}
   \label{fig:l0_n50_D4}
 \endminipage\hfill
 \caption{Variation of the Rényi entropy power, $\mathcal{N}_{p}[\rho]$, with respect to $p$ for the Rydberg oscillator state ($n=50$, $l=0$) of the $D$-dimensional harmonic oscillator with $D=2 (\blacksquare)$ and $D=4(\bigodot)$.}
 \label{fig:renpow}
 \end{figure}

First, in Figure~\ref{fig:renpow} we study the variation of the Rényi entropy power, $\mathcal{N}_{p}[\rho]$, with respect to the order $p$ for the Rydberg oscillator state ($n=50, l=0$) with $D=2(\blacksquare)$ and $D=4(\bigodot)$. We observe that in both cases, the Rényi entropy power decreases monotonically as the order $p$ is increasing; in fact, this behavior holds for any dimensionality $D>1$. Moreover it is very fast, indicating that the quantities with lowest orders (particularly the case $p=2$, closely related with the disequilibrium) are most significant for the quantification of the spreading of the electron distribution of the system.\\

\begin{figure}[H]
\minipage{0.50\textwidth}
\includegraphics[width=\linewidth]{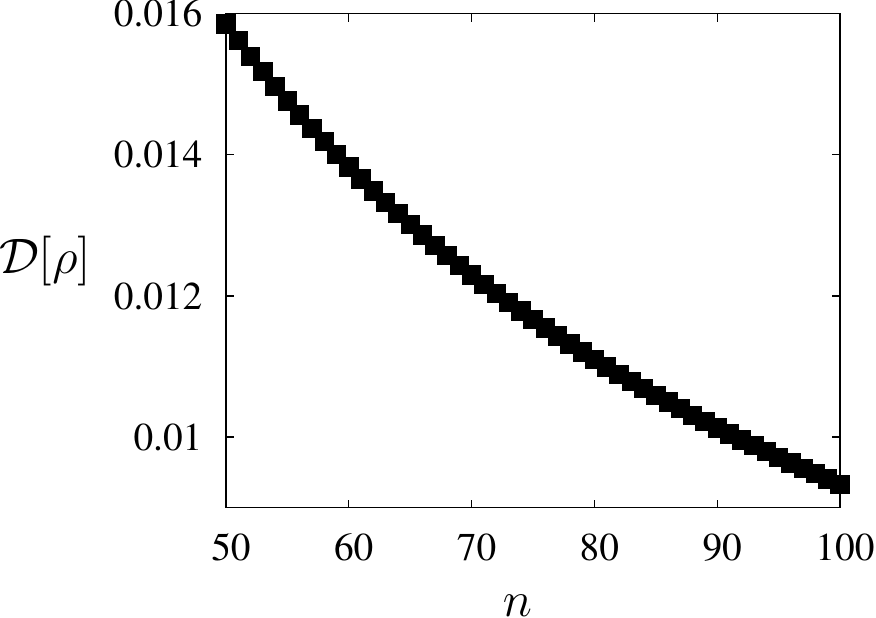}
\caption{Variation of the disequilibrium $\mathcal{D}[\rho]$ with respect to $n$ for the Rydberg oscillator ($ns$)-states of a two-dimensional harmonic oscillator. So, this is the case ($l=0$, $p=2$, $D=2$).}
\label{fig:l0_p2_D2}
\vspace{4mm}
\endminipage\hfill
\minipage{0.50\textwidth}
  \includegraphics[width=\linewidth]{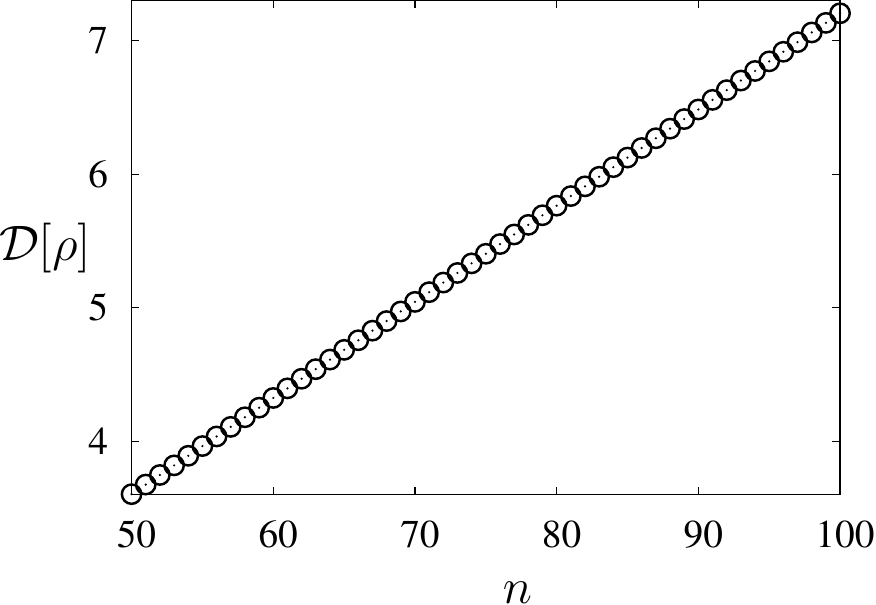}
  \caption{Variation of disequilibrium $\mathcal{D}[\rho]$ with respect to $n$ for the Rydberg oscillator $(ns$)-states of a six-dimensional harmonic oscillator. So, this is the case ($p=2$, $l=0$, $D=6$).}  
 \label{fig:l0_p2_D6}
\endminipage\hfill
%\caption{Variation of disequilibrium, $\mathcal{D}[\rho]$, with respect to $n$ for the Rydberg oscillator state for $l=0$, $p=2$ with $D=3 (\bullet)$ and $D=6(\bigodot)$.}
\label{fig:diseq}
\end{figure}

\begin{figure}[H]
  \minipage{0.50\textwidth}
  \includegraphics[width=\linewidth]{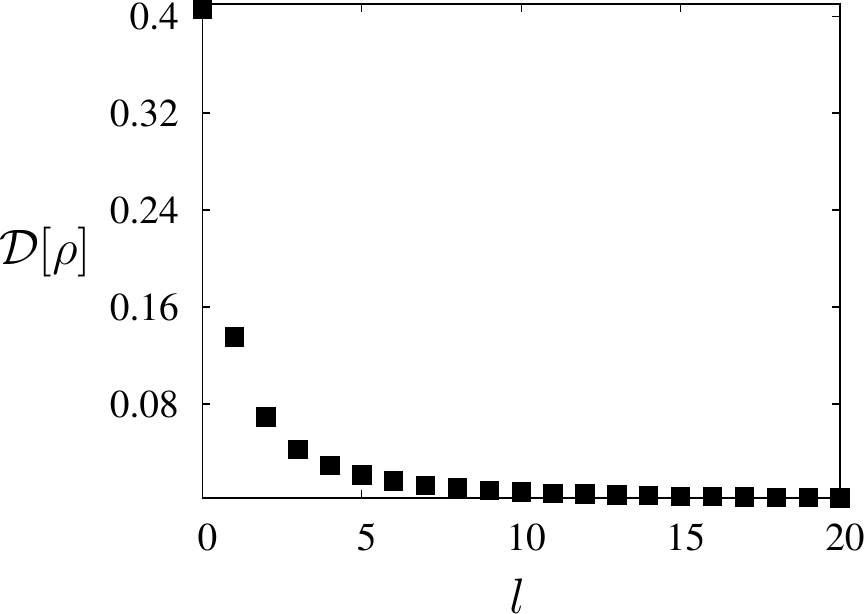}
\caption{Variation of the disequilibrium $\mathcal{D}[\rho]$ with respect to $l$ for the Rydberg oscillator state with $n=50$ of the four-dimensional harmonic oscillator %(\blacksquare) 
. So, this is the case ($p=2$, $n=50$, $D=4$).}
 \label{fig:p2_n50_D4}
 \vspace{4mm}
  \endminipage\hfill
  \minipage{0.50\textwidth}
    \includegraphics[width=\linewidth]{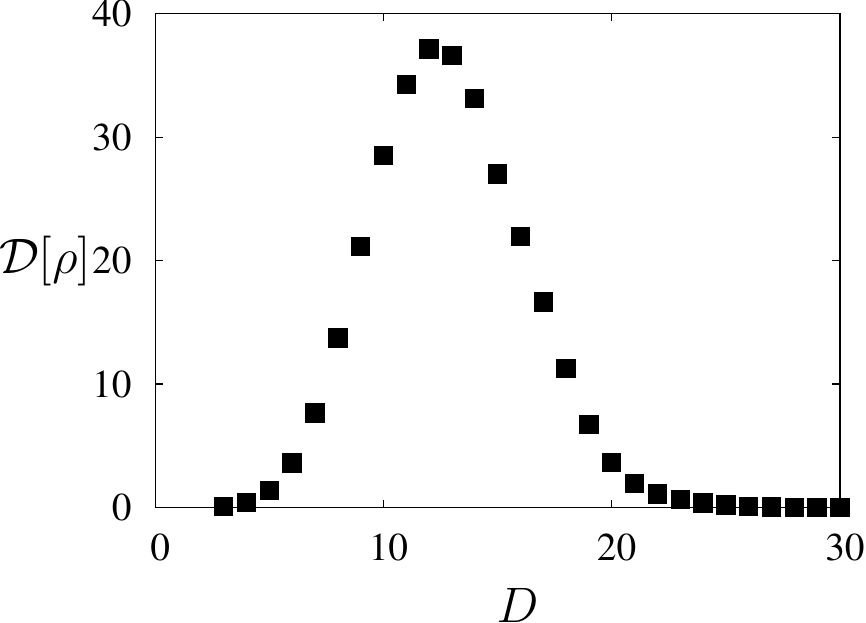}
    \caption{Variation of the disequilibrium $\mathcal{D}[\rho]$ with respect to the dimensionality $D$ for the Rydberg oscillator state ($n=50$, $l=0$) of the $D$-dimensional harmonic oscillator %(-)$
    . So, this is the case ($p=2$, $n=50$, $l=0$).}
   \label{fig:l0_p2_n50}
  \endminipage\hfill
%  \caption{Variation of the disequilibrium, $\mathcal{D}[\rho]$, for the Rydberg oscillator state for $p=4$, $n=50$ and $D=2 (\blacksquare)$ and $l=0$, $p=2$ and $n=50(-)$, respectively.}
  \label{fig:renpow_diseq}
  \end{figure}

Second, we explore the dependence of the pth-order Rényi quantities of the Rydberg-state region in terms of the principal hyperquantum number $n$ when ($l, p, D$) are fixed. To exemplify it, we will examine the case $p=2$ for the Rydberg ($n, l=0) \equiv (ns$)-states  of the three-dimensional oscillator. From (\ref{eq:entropmom2}), (\ref{eq:rhonl}), (\ref{eq:c1.2}), (\ref{eq:renyiasymp}) and Theorem 1, one has that for $p=2$ and $D=3$, the second-order Rényi entropy and the disequilibrium (the inverse of the Rényi entropy power) of the Rydberg state ($n,l$) are given by
\begin{eqnarray}
\label{eq:renyip2D3}
R_{2}[\rho] &=& -\ln\left[2\lambda^{\frac{3}{2}}C_{B}\left(l+\frac{1}{2},-\frac{1}{2},2 \right) n^{-\frac{1}{2}}\right]\,(1+\bar{\bar{o}}(1)), \nonumber \\
\mathcal{D}[\rho] &=& W_{2}[\rho] = \mathcal{N}_{2}[\rho]^{-1} \nonumber \\
 &=& \left[2\lambda^{\frac{3}{2}}C_{B}\left(l+\frac{1}{2},-\frac{1}{2},2 \right) n^{-\frac{1}{2}}\right]\,(1+\bar{\bar{o}}(1))
\end{eqnarray}
since the disequilibrium (or average density of the distribution $\rho$) is defined as $\mathcal{D}[\rho] := \int \rho(x)^{2}\, dx$. Moreover, for the $(ns)$-states the $C_{B}$-constant given in (\ref{4}) can be explicitly calculated, so that the second-order Rényi entropy and the disequilibrium of the Rydberg $(ns)$-states of the three-dimensional harmonic oscillator has the following behavior
\begin{eqnarray}
\label{eq:renyip2D3l0}
%R_{2}[\rho] &=&- \ln\left[\frac{2\lambda^{\frac{3}{2}}}{\pi}n^{-\frac{1}{2}}\right] \,(1+\bar{\bar{o}}(1))\nonumber\\
R_{2}[\rho] &=& \left[-\ln\left(\frac{2\lambda^{\frac{3}{2}}}{\pi}\right)+\frac{1}{2}\ln n\right] \,(1+\bar{\bar{o}}(1)),\nonumber \\
\mathcal{D}[\rho] &=& \left[\frac{2\lambda^{\frac{3}{2}}}{\pi}n^{-\frac{1}{2}}\right] \,(1+\bar{\bar{o}}(1)),
\end{eqnarray}   
respectively. The case ($l=0, p=2, D=2$) as well as the case ($l=0, p=2, D=6$) are plotted in Figures \ref{fig:l0_p2_D2}-\ref{fig:l0_p2_D6}, which gives the variation of the disequilibrium, $\mathcal{D}[\rho]$, with respect to $n$ for the Rydberg oscillator ($ns$)-states of the two- and six-dimensional harmonic oscillator, respectively. We observe that the behavior with respect to $n$ for the disequilibrium of these states has a decreasing (increasing) character in the two (six)-dimensional oscillator. On the other hand, one can realize from (\ref{eq:renyiasymp}) and Theorem 1 (see the last lines of the comments to this theorem) that the disequilibrium for the case ($l=0, p=2, D=4$) has the constant value $0.4053$. So that, most interesting, we find the following phenomenon: the disequilibrium of the Rydberg ($ns$)-states of $D$-dimensional oscillator decreases (increases) as a function of the principal hyperquantum number $n$ when the dimensionlity $D$ is less (bigger) than $4$, and it becomes constant when $D=4$. In fact we should not be surprised that the disequilibrium as a function of $n$ changes when the spatial dimensionality is varying. This also happens for all physical properties of a quantum system at different spatial dimensionalities, since the physical solutions of their corresponding wave equations (e.g., Schrödinger) are so different (see e.g., \cite{dong}). The novelty is that the character of the disequilibrium behavior as a function of $n$ changes so much, pointing out the existence of a critical dimensionality at which it is constant.\\
In Figure~\ref{fig:p2_n50_D4}, we illustrate the variation of the disequilibrium, $\mathcal{D}[\rho]$, as a function of $l$ for the Rydberg states ($n=50, l$) of the four-dimensional harmonic oscillator. We observe that its behavior is monotonically decreasing when $l$ is increasing. In fact this property holds for $D\ge 2$. Then, it is interesting to point out that the electron distribution of the $D$-dimensional oscillator, within the region of the Rydberg $s$-states, becomes closer to equiprobability when $l$ is increasing, approaching what one would expect classically. Moreover, this trend is slightly moderated for Rydberg states other than $s$-states.\\
Third, finally, let us illustrate the behavior of the Rényi entropy power, $\mathcal{N}_{p}[\rho]$, of the Rydberg oscillator states as a function of the dimensionality $D$. We do that in Figure~\ref{fig:l0_p2_n50} for the disequilibrium $\mathcal{D}[\rho] = \mathcal{N}_{2}[\rho]^{-1}$ of the Rydberg state ($l=0, p=2, n=50$) of the oscillator with various integer values of the dimensionality $D$. We observe that the disequilibrium has a \textit{quasi-Gaussian} form when $D$ is increasing, so that finally it vanishes for a given, sufficiently large value of $D$. Most interesting is that the maximum of the Rényi entropy power is located at $D=12$, which surprisingly corresponds to the universe dimensionality predicted by certain string theories \cite{bars}. Nevertheless, we should point out that for higher Rydberg states the maximum of the disequilibrium is located at larger dimensionalities. This indicates that the nearer the classical limit is, the larger is the dimensionality required for the disequilibrium (i.e., separation from equiprobability) to reach its maximum.

\section{Conclusions}

The macroscopic properties of a quantum many-particle system essentially depend on the spreading of its quantum-mechanical Born one-particle distribution $\rho(\vec{r})$, as proved by the functional-density theory. This spreading can be completetely described by the knowledge of the moments $W_{p}[\rho]$ of $\rho(\vec{r})$ or by some closely related information-theoretic quantities, the Rényi entropies $R_{p}[\rho]$, which often describe some fundamental properties of the system and/or are experimentally observable. These quantities, however, cannot be analytically accessible, even not for the simplest harmonic systems unless we consider the ground state and the first few lowest-lying excited states. In 2012 the Shannon entropy, which corresponds to  the limiting case $p\rightarrow 1$ of the Rényi entropy, was determined for the highest-lying (Rydberg) states of the one-dimensional harmonic oscillator \cite{ADST} whose wavefunctions are controlled by Hermite polynomials.\\

In this paper we extend this result in a two-fold way: we determine in an analytical way the Rényi entropies of all orders for the Rydberg states of a $D$-dimensional harmonic oscillator, whose wavefunctions are known to be controlled by Laguerre polynomials. To do that we first realize that the Rényi entropies can be explicitly expressed in terms of the $\mathcal{L}_{p}$-norms of the Laguerre polynomials, and then we develop a method to analytically calculate the leading term of the asymptotics of these norms when the polynomial degree is very high.\\

 Later, a number of physical results are found. First, for a given Rydberg state the Rényi entropy has a very fast decreasing behavior as the parameter order is increasing, indicating that the Rényi entropies with lowest orders are most significant. Then, for illustration, we study in detail the second-order Rényi entropy (i.e., the disequilibrium) of the system, which quantifies the separation of the electron distribution from equiprobability. It is found that it  has a \textit{bell-like quasi-Gaussian} behavior in terms of $D$, its maximum being located at $D=12$ which is the universe dimensionality predicted by certain string theories \cite{bars}. Let us here comment that geometrical quantities associated with $D$-dimensional hyperspheres (such as surface area) also exhibit this kind of behavior (with the corresponding bell-like function centered around a different $D$-value). This suggests that the behavior of the disequilibrium may have a geometrical origin in terms of basic properties of hyperspheres. Moreover, the disequilibrium of the Rydberg ($ns$)-states of $D$-dimensional oscillator decreases (increases) as a function of the principal hyperquantum number $n$ when the dimensionlity $D$ is less (bigger) than $4$, and it becomes constant when $D=4$. Needless to say that much more efforts have to be done before making exotic statements.\\
  
Finally, these results are potentially useful in the study of entropic uncertainty relations. Moreover, they might also be relevant in connection with quantitative entanglement indicators. Rényi entropies have been recently used for this purpose (see e.g., \cite{zander}). We believe that the analytical technology here developed could be useful in relation to entanglement-like studies in quantum information. 

\section*{Acknowledgments}
The work of A.I. Aptekarev and D.N.Tulyakov was supported by the grant of Russian Science Foundation (project 14­21­00025). The work of J.S. Dehesa was partially supported by the Projects
FQM-7276 and FQM-207 of the Junta de Andaluc\'ia and the MINECO grants FIS2014-54497P and FIS2014-59311P. The work of I. V. Toranzo was supported by the program FPU of MINECO. Useful numerical discussions with Dr. Pablo Sánchez-Moreno are acknowledged.

\end{document}